**EMERGING TECHNOLOGIES**

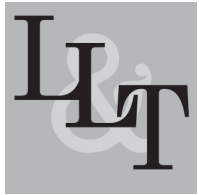

# Distributed agency in second language learning and teaching through generative AI

*Robert Godwin-Jones, Virginia Commonwealth University*

## Abstract

*Generative AI offers significant opportunities for language learning. Tools like ChatGPT provide second language practice through chats in written or voice formats, with the learner specifying through prompts conversational parameters. AI can be instructed to give corrective feedback and create practice exercises. Using AI, instructors can build learning and assessment materials in a variety of media. Generative AI provides affordances for both autonomous and instructed learning. In addition, AI is poised to enhance dramatically the usefulness of immersive technologies. For both learners and teachers, it is important to understand the limitations of AI systems that arise from their statistical model of human language, which constrains their capacity for dealing with sociocultural aspects of language use. Additionally, there are ethical concerns over how AI systems are created and deployed, as well as practical constraints in their use, especially for less privileged populations. Nevertheless, the power and versatility of AI tools are likely to turn them into constant companions in many people's lives, creating a close connection that goes beyond simple tool use. Ecological theories such as sociomaterialism are helpful in examining the shared agency that develops through close user-AI interactions, as are the perspectives on human-tool relationships from Indigenous cultures.*

## Introduction

The arrival of generative AI (artificial intelligence) is reshaping our world in any number of ways. As AI is integrated into ever more tools and services, it is becoming ubiquitous in desktop, mobile, and wearable devices. Workplace environments of all kinds are seeing increased AI automation. Change is already evident in jobs and tasks that involve routinized writing, as *ChatGPT* and similar tools have shown remarkable abilities in areas such as summarizing, synthesizing, and authoring in conformity to specific genres – from weather forecasts and argumentative essays to storytelling and computer coding. AI integration will bring greater efficiency in many areas and at the same time considerable social disruption. Some lower-level jobs are likely to be handed off to AI. Pundits, politicians, and philosophers worldwide are grappling with the practical and ethical repercussions of an AI world, as well as what powerful AI may mean for the future of humanity.

The extent of societal change through generative AI is not yet clear, even as systems improve continuously and become more frequently integrated into existing digital services of all kinds. The "AI" moniker is now being attached to products and services of all kinds, similar to the hype around web addresses in the 1990's or surrounding mobile apps in the early smartphone days. AI frenzy has extended to educational settings. In the short time it has become publicly accessible, generative AI has already wrought significant changes in education. That shift is most evident in writing assignments and assessments but affects all disciplines. Ways to deal with student use of *ChatGPT* have ranged from banning its use to fully integrating it into coursework. Efforts to identify AI-created artifacts through detection tools or watermarks have so far proven ineffectual.

Second language (L2) learning and teaching is not exempt from the AI wave. Students have discovered that *ChatGPT* is quite adept at writing coherently in a variety of genres, providing substantive and well-organized answers to questions, and able to communicate in multiple languages. It is also capable of correcting grammar, improving writing style, and offering high-quality translation. Teachers have discovered AI's ability to create lesson plans; write texts or learning activities at any given level of proficiency; and provide assessments of student work. Since November 2022, a multitude of blog posts, YouTube videos, and conference papers have explored the capabilities of *ChatGPT* for language learning. Peer-reviewed articles are appearing as well (Bonner et al., 2023; Kohnke et al., 2023; Moorhouse, 2024). Educational institutions around the world are issuing guidelines for AI, while professional language teacher organizations are offering workshops on integration of AI into language learning and teaching.

The buzz around AI is not misplaced. It is my contention in this column that the uses of generative AI we have seen as yet are but the tip of the iceberg, and that multiple profound changes are on the way as generative AI tools become ubiquitous (at least in privileged communities). As AI takes on roles providing, on the one hand help in improving writing skills, and on the other hand informal conversation in the L2, the nature of SLA (second language acquisition) will shift, as will the role of teachers in the learning process. I argue here that the appropriate initial response to AI in instructed SLA settings should be the development in learners (and in teachers) of critical AI literacy—an understanding of the nature of generative AI and of the power dynamics its use represents (Dooly & Darvin, 2022). That will include gaining insight into the ethical concerns surrounding AI production as well as the practical barriers to its use.

To be able to integrate AI effectively in instructional settings, teachers and learners will need to have an appreciation of AI's evolving relationship to humans and their environment. In that process, ecological frameworks such as sociomaterialism can be helpful to illuminate the emergent, entangled agency — shared between human users and AI — that typify AI use for language learning. Teachers will want AI use to reflect the need for equitable multilingualism in SLA as well as for an awareness of social justice issues. Calling on indigenous, non-Western ontologies can be helpful in understanding the need for an inclusive perspective on linguistic plurality as well as for an acceptance of symmetry in control and action between humans and natural or artificial entities. The column concludes with speculation on the future of language learning in an AI-infused world, with particular attention to the role that AI will play in immersive learning environments.

Our examination of AI in language learning is organized around a succession of perspectives on a central issue, namely, where does *agency* reside in the AI-learner-teacher relationship, who controls the interactions. Agency is seen here as "the socio-cultural mediated capacity to act" (Han & Reinhardt, 2022, p. 989), a property conceived therefore as relational and contingent, emergent rather than inherent. Agentive capacity resides in humans, but also in objects (Guerrettaz et al, 2021), including digital tools. Following that dynamic, many issues arise for AI use in language learning and teaching:

- What form can human agency take in dealing with the extraordinary power of AI?
- How can AI literacy enhance learner and teacher agency?
- Which forms of AI integration are becoming available and how do they affect user control?
- What does shared agency in L2 writing look like?
- How can we frame theoretically the human-tool dynamic in AI use?
- Do the ethical issues and practical problems inherent in the creation and use of AI systems weigh more heavily than the gains in learner agency?
- Finally, how does one possible future for language learning reside in an integration of AI into immersive technologies, with a resultant merged human-AI agency?

## AI Asserts Control: "I'm sorry, Dave, I'm afraid I can't do that."

HAL, the computer in *2001, A space odyssey,* trying to kill astronaut Dave by refusing to open the ship's airlock, represents the doomsday scenario of rogue supercomputers and killer robots familiar from science fiction, but also evoked in a statement by leading AI researchers that generative AI may pose an "existential" threat to humankind (Roose, 2023). The notion that AI is capable of independent actions, exhibiting something like human consciousness, does not seem so far-fetched when one considers the assertion by a Google engineer that evidence from chats he conducted with Google's *LaMDA* had convinced him that the AI system had become sentient (Grant & Metz, 2022). Although the engineer was fired, his estimation of the ability of a generative AI system to emulate convincingly something akin to human interaction is likely not far off from what many users of *ChatGPT* have experienced themselves. That impression goes beyond the well-known tendency of humans to attribute human-like qualities to objects able to communicate (anthropomorphism; Holtgraves et al., 2007). In fact, generative AI systems and emerging AI agents (such as Pi, the "personal AI assistant" or ElliQ, the chatty robot for the elderly) are far from passive, robotic partners; they can act proactively, able to initiate conversational turns based on prior conversations.

That does not mean that we have reached the "singularity" of super intelligent systems predicted by Ray Kurzweil (2005) or even the level of general AI (GAI) that emulates high quality human-like performance in virtually all domains of human activity. Whether genuine GAI will ever be possible is an open question and certainly not a prospect universally embraced. The case for assuming an ever-expanding list of AI capabilities can be made by considering the emergent properties of systems like *ChatGPT*. AI engineers have been surprised by some of the functionality users uncovered, which had not been anticipated (Kosinski, 2023). Given the constant improvement across subsequent *ChatGPT* models (versions 3 to 3.5 to 4 to 4o), additional capabilities and improved performance in chat interactions can be expected.

GAI may not be on the immediate horizon, but in particular domains there is ample evidence of high-level performance (Morris et al., 2024; Riedl, 2023). Machine learning and enhanced media abilities in AI systems will move them to ever higher levels of ability in language output (Patil et al., 2024), reasoning capabilities (Anderson, 2024), and multimodal integration (Metz, 2024). Advances in machine learning technology such as reinforcement learning and genetic algorithms point to the potential of AI to reconfigure itself for improved performance without human intervention (Dattathrani & De', 2023). Self-rewarding models in generative AI enable an iterative process that enhances the ability of AI systems to self-organize and to improve methods of learning (de Gregorio, 2024), reducing or possibly eliminating the need for human feedback training.

Such developments obviously raise a host of issues in terms of AI's role in human society. There are already widespread concerns over the potential of AI to replace humans in a many jobs. In addition to the economic impact — old jobs departing, new ones arising — the social dimension of more powerful AI raises questions of what it means for AI agents to move from assistants to companions. To what extent will we rely on systems that have developed autonomously? Can they be trusted to have our interests foremost? What level of influence will they have in our day-to-day behaviors and interactions? These questions have a particular relevance in education, as the presence of adaptive AI tutors, along with other AI-powered services, would seem to necessitate a rethinking of education generally and particularly of the role of teachers.

For language learning, AI promises breakthrough advances for informal, incidental language learning through the availability of written and voice chatbots, as well as for structured learning through individualized tutors, fulfilling the long-term dream of tutorial CALL (computer-assisted language learning). In fact, the versatility available in setting parameters for AI interactions through prompts (open-ended exchange, formal tutoring, friendly advising, etc.) blurs the artificial division between implicit and explicit SLA. From the perspective of language learning, it is irrelevant whether our conversation partner is sentient or not. The fact that we can even ask such a question demonstrates AI chatbots' extraordinary effectiveness as believable conversationalists, and therefore extreme usefulness as human-like, always available, non-judgmental chat partners. Through AI, the changes in the relationship of learner-tool-environment are likely to be profound, representing dynamically shifted ecosystems for language education. To understand how such a structural change may occur, what it may mean for learners and teachers, and how such a system may play out, we will need to go beyond viewing AI simply as another technology tool (Warschauer et al., 2023), but instead consider holistically AI's impact on all stakeholders.

Helpful in that process will be to turn to theories that examine the relationship of humans and nonhumans in larger social and cultural contexts, namely ecological frameworks. A starting point is Leo van Lier's work on *affordances*. Affordances are one way to envision how humans interact with the material world. Van Lier (2010) viewed affordances as relationships of possibilities that were dependent on situated action between a learner and tools/artifacts in a specific context. That perspective was taken up by later ecological theories that consider affordances as a relational property (Jarzabkowski & Pinch, 2013). Properties of an object acting as an affordance are not static and fixed but emerge through their relationship with the learner.

Atkinson et al. (2018) expanded on the concept of affordances in their insistence on the importance of an *alignment* of resources to learners' "mindbodyworld," emphasizing the agentive presence of physicality and environment (p. 474). From this perspective, learner engagement may be "mindful and intentional, or not, or anywhere in between" (Atkinson et al., 2018, p. 488). In addition, from an ecological perspective, "any object may afford multiple possibilities that are beyond those purposes for which it is designed" (Jarzabkowski & Pinch, 2013, p. 582). In that way, objects/tools can be repurposed through human actions. That applies to technology tools. In computer-mediated communication and social media, language learners use the affordances of those tools to make contributions, write posts, react to others, and in the process make changes to the systems themselves. Similarly, in AI, users' input changes the system, as every prompt becomes system data and leads to further machine learning. There is a two-way process at play.

The extent of user control varies among digital tools. Interactions in social media (*Facebook, Instagram, TikTok*) are governed by algorithms that reflect previous user choices, but also by commercially determined factors such as advertising/marketing considerations or even the desire to rank extreme views more highly so as to foment buzz and generate greater money-making online activity (Dooly & Darvin, 2022). Generative AI tools like *ChatGPT* have built-in biases and control mechanisms (as discussed below), but they do provide a higher degree of user control than do algorithmically driven social media. AI chatbots enable the user to decide, within limits, on both content and conversational parameters. Users determine the flow of the conversation and are able to modify output through subsequent prompts.

Seen from an ecological perspective, the ability to affect change purposefully — that is, agency — resides in both the user and in the AI system: "Through their semiotic symbol processing capability, they [information systems] are also digital actors capable of performing social action on behalf of humans and organisations" (Ågerfalk, 2020, p. 2). The capability of AI to act autonomously on behalf of humans does not presuppose consciousness, but it does represent a contingent form of agency. That agency is reciprocal but not necessarily symmetrical, as AI use needs to be constrained by rights, responsibilities, and ethics (Dattathrani & De', 2023). This does not translate into the traditional, enlightenment-inherited view of humans as standalone, rational individuals exercising unique control over the material world. Instead, there is a "shifting entanglement of humans and nonhumans" (Hasse, 2019, p. 360), necessitating a nuanced, contextually determined understanding of agency when it comes to humans and AI.

The automatic assumption for many teachers is that AI may replace learner agency, that is, that it will be used by students in a copy and paste fashion to complete assignments. That parallels suppositions language teachers may have had for some time concerning the use of machine translation (MT). Actual studies of student MT use have demonstrated the fallacy of that view, in that students tend to use MT selectively, for looking up words or phrases or checking individual sentences (Hellmich & Vinall, 2021; Jolley & Maimone, 2022; Vinall & Hellmich, 2022). While it is too early to know if AI use will follow that pattern, emerging studies, discussed below, show in fact that exchanges between AI and learners are not typically one-way but circular. Through that process, agency can be seen as emergent and distributed, arising from iterative interactions between human user and AI. At least for now, in interacting with AI, humans are still "in the loop" (Mosqueira-Rey et al., 2023), or to cite Floridi (2023b), "*Terminator* is not coming" (p. 208).

## Through Critical AI Literacy, Learners and Teachers Gain Agency

AI use for learners and teachers will be optimized through achieving a basic understanding of how generative AI systems work. Informed consumers of AI will be aware of both capabilities and limitations of AI, leading to the development of "calibrated trust" (Ranalli, 2021, p. 14) in AI systems. *ChatGPT* and other tools based on generative AI have become expert language users in a way that is very different from how humans learn and use language. Early efforts in AI attempted to instill in computers the linguistic properties of a given language, that is its syntax, morphology, phonology, and semiotics. That rule-based training was accompanied with efforts to teach AI systems fundamental aspects of how human society and the physical world operate. A famous example is the CYC project (Lenat, 1995). That approach had partial success, particularly within limited domains of human activity in which predictable human behavior and typical language patterns could be anticipated and partly pre-programmed.

Generative AI systems use a quite different approach to developing language abilities, one that has proven to be phenomenally successful. Through analyzing immense volumes of language data (collected by scraping indiscriminately digital text sources) and using machine learning techniques based on the "transformer" architecture model (Vaswani et al., 2017), the system discovers on its own how human language works. For the "large language models" (LLMs) emerging from this process, language is a statistical problem, centered around word use frequency and contextual use/reuse/variation. By identifying patterns and regularities in language (through translating chunks of language – "tokens" – into mathematical symbols – "vectors" – then combining vectors into "parameters"), LLMs can predict the next word in a text sequence, which then is extrapolated to the ability to write phrases, sentences, paragraphs, lengthy connected discourse. The complex algorithms, searching and sifting through multiple layers of an artificial neural network accessed simultaneously ("deep learning"), work efficiently and swiftly, using vast arrays of powerful computers. Interestingly, the probabilistic approach LLMs use does not automatically select the first item identified (i.e., the most commonly used word for that context) but rather may pick a word further down the frequency list, depending on the context (derived from the use of the "attention" process of the transformer architecture). That results in texts which avoid mechanical banality and approach human naturalness and fluency in writing style. Wolfram (2023) provides a detailed account of how LLM's like *ChatGPT* work.

Generative AI relies on a mathematical rather than a linguistic model of language. LLMs work so well at generating language because their training is based not on fixed rules but on actual language use (namely, its training corpus). It has been argued that the language model represented in generative AI "is a clear victory for statistical learning theories of language" (Piantadosi, 2023, p. 18). LLM's validate a usage-based language model, based not on grammar rules but on the use, reuse, and transformation of multi-word units such as collocations, idiomatic phrases, or syntactical patterns (frame and slot structures, for example). In this view, language "consists of regular overlapping sequences in the form of chunks that are processed, stored, and retrieved as wholes rather than being constructed bottom-up from grammar 'rules' as traditionally thought" (Boulton, 2017, p. 189). Given that model of language, it is not surprising that Noam Chomsky (2023), championing as he does an innate, human language ability, finds little to admire in large language models.

Language output from *ChatGPT* and from other LLM-based chatbots is coherent, grammatically accurate, idiomatically valid, with appropriate transitions among sentences and paragraphs. The generated text conveys the impression that it could have been written by a human being. The quality and speed of output tend to instill confidence in the generated texts, as if coming from a reliable source. The text is presented as if it were a series of objective, factual statements, no matter whether the information it contains is true or not. Paradoxically, the highly sophisticated system that creates these extraordinary texts has no real understanding of what they mean. The texts are in a sense simply repackaging (in elegant form) of the data the system has been fed, achieved through a complex mathematical process of analysis and reconstruction. The absence of true understanding of language and its meaning can lead, at times, to the production of inaccurate statements and inappropriate texts. Early users of *ChatGPT* reported disturbing examples of the system's "hallucinations," making misplaced, even offensive comments and judgments about the user.

The authoritative sounding voice of generative AI has no real-world knowledge or experience to draw on. AI systems have not been socialized into language use, as are humans, and that has consequences for their understanding and use of socially and culturally determined language, in other words, its *pragmatic* competence. AI has been able to mimic humans in terms of knowing linguistic patterns and behaviors expected in given social situations (i.e., sociopragmatics). That involves, for example, the use of politeness formulas, the degree of directness in making requests, or the expected expressions to use in apologizing (in a particular cultural context). Where AI has been less successful is in the ability to adjust output based on context and interlocutor characteristics (i.e., pragmalinguistics). Individual speakers may have knowledge of expected norms and behaviors in a given culture, but for a variety of reasons — personal, interpersonal, political — may choose in a specific context to modify or reject their use (see Garcia-Pastor, 2020 and McConachy, 2019 for examples).

Pragmatics is a difficult area to navigate, even for humans, given the sensitivity and emotional weight of pragmatic language choices. We negotiate nuances of socially and culturally framed language through inference, intuition, and human lived experience, and even so we can sometimes get it wrong. AI systems don't have those resources to draw on in interpreting meaning and adjusting language output accordingly. A study of the use of *ChatGPT* in applying pragmatic language in Chinese reports that the system could use pragmatic language appropriately but had difficulty finding the right language to use in contexts that involved speaker subjectivity (Su & Goslar, 2023). On the other hand, the researchers found that in subsequent versions of *ChatGPT* — from week-to-week new releases — performance improved; it is likely that *ChatGPT* was learning from scenarios the researchers fed repeatedly in their experimentation. More studies of AI and language pragmatics would be welcome, given how essential pragmatic appropriateness is to effective human communication.

Humans develop in childhood through social interactions the ability to judge the state of mind, intentions, and feelings of those around us. We do that through language, but also through reading body language, gestures, facial expressions, and paralanguage (tone, intonation). Humans can recognize and understand actual meaning in sarcasm, implicature, or emotional masking. In textual exchanges with an AI, the nonverbal dimensions of language are absent, thus robbing the system of valuable messages as to meaning and intention. It is increasingly recognized in applied linguistics that human communication is not just verbal, but that meanings are embodied, embedded, enacted, extended (4E cognition theory; Ellis, 2019; Witte, 2023). This framing of human cognition takes into account the myriad semiotic resources that make up language exchanges (Godwin-Jones, 2023a). Semiotic systems enabling human communication constitute more than verbal language and have meaning beyond transactional exchanges (Godwin-Jones, 2019a). Often real human conversations convey emotional messaging or serve the purpose of making/maintaining social connections.

In addition to the conception of human communication as a broad set of semiotic resources and repertoires, another currently accepted theory in applied linguistics that is problematic for AI is the reality of *translanguaging* (García & Kleifgen, 2020; Wei, 2018), the continual co-presence in our minds of all the languages we speak. That leads away from a conventional view of human communication, often found in L2 classrooms (and in CALL studies), as being overwhelmingly verbal and essentially monolingual. AI systems can deal with multiple languages, but they do not have the lived experience of bilinguals in code switching, which is subject to multiple variables including setting, speakers' dynamic identity/intent positioning, and myriad sociocultural factors (Kramsch, 1998).

Ortega (2019) summarizes current SLA theories on the nature of language in this way:

> Humans make meaning by assembling linguistic signs but also by pooling language (and all their languages) together with whatever other bits of semiotic repertoire they have, to the point that meaning making is always multisensory, multimodal, and always involving much more than language (pp. 290-291).

This complicates the task for AI systems' capacity to communicate the way that humans do, lacking as they are in sensory organs, physical signaling, and sensitivity to shifting sociocultural cues and clues. Yet even without the ability to observe humans in the act of communicating, AI systems have shown some capacity for reading and responding to subjective cues. In fact, there is evidence that AI has been able to develop some degree of a "theory of mind", the ability to judge the subjective state of an interlocutor (Kosinski, 2023). That capability could be enhanced through computer vision technology that reads and interprets facial expressions. Likewise, in spoken exchanges, the capacity to decipher tone of voice and intent through intonation or other means is likely machine learnable. GPT 4o (released May 2024) already demonstrates a remarkable ability to read a speaker's emotional state and to respond in a wide emotional range (Lohrbeer, 2024). At the same time, even vision-capable and synthetically empathetic AI robots will lack communicative human experiences, especially the negotiation of common ground between speakers. That limits the ability to share emotions or create joint experiences. In the view of Potamianos (2022), AI systems have, "the social cognitive capabilities of a 2-year-old" (para 9). The social IQ of AI is bound to improve, yet its lack of human experiential knowledge will continue to be a limiting factor.

For both language learners and teachers, an understanding of AI's language skills and limitations is important in being realistic about expectations. That initial step towards AI literacy needs to be followed by development of knowledge in how to use AI systems effectively and ethically. A framework for that process is outlined in Tseng and Warschauer (2023) and Warschauer et al. (2023). AI can be useful in L2 text creation for brainstorming writing ideas, for suggesting ways to model texts to particular styles or genres, and for transforming text to address particular audiences or levels of language proficiency. Instructional prompts to the AI can specify particular kinds or levels of feedback to provide on learner written texts. In instructed settings, the use of AI for different tasks can be demonstrated and modeled, with reflection and discussion of teacher and student experiences. McKnight (2021) presents multiple classroom activities integrating AI, many of which emphasize careful, collaborative reflection on AI use.

A critical perspective on AI also involves an understanding of the differential power relationships inherent in AI use. Access to AI resides in the hands of large tech companies (OpenAI, Meta, Microsoft, Google), who decide issues of cost, geographical distribution, delivery (Web, mobile app), code availability, and end user conditions. AI companies with the widest user base (mostly US-based) are commercial, for-profit entities, that answer to shareholders, boards of directors, and, to a lesser extent, regulatory agencies. Users can opt to use open source LLMs, which are widely available (as well as lightweight versions installable on personal devices), but the most advanced functionality is likely to require use of commercial, cloud-based versions, given the enormous development and deployment costs of LLMs. AI services are available through different devices and modalities, with the likelihood of expansion into wearable and embedded devices. The mode of access to digital tools and services affects both functionality and the degree/nature of user control (Darvin, 2023), so studies of differentiated user experiences and learning outcomes depending on AI access mode will be valuable.

Language teachers integrating AI into instruction will want to address general issues of AI development and fairness, as discussed below. In implementing AI functions in learning tasks, it is important to stress the critical evaluation of AI output. Peer collaboration and teacher modeling can contribute to L2 learner's ability to understand how to integrate AI output into personal narratives that showcase individual voice and creativity. This is a crucial skill that is useful outside of academic settings: "Through teaching students how they can tangibly use these tools in their own learning, educators can create strong foundations for their students' immediate learning and long-term use of AI-based tools in educational and professional contexts" (Tseng & Warschauer, 2023, p. 260). Rather than banning or discouraging AI use, teachers need to help students deal in a balanced and critically informed way to use AI to optimally represent their own ideas and identities. AI literacy should be critical (i.e., understanding power dynamics), but also agentive, indexing the capacity and will to act. That is important for academic success, but also for life beyond the classroom.

## Varied Reach and Power of AI: Narrow and General Use Implementations

AI has been in use for some time in tools and services that target a particular area of language use or learning, such as machine translation (*Google Translate*), intelligent text editors (*Grammarly*), or automated writing evaluation (*Criterion*). These examples of *narrow AI* (Schmidt & Strasser, 2022) represent important tools for language teachers and learners (see Godwin-Jones, 2022, for an overview). However, the capabilities of general-purpose AI systems promise to take the place of many of their functions. Rather than being trained to target a particular domain, LLM-based systems like *ChatGPT* are designed to function broadly within virtually any sphere of human activity. Additionally, general AI services from OpenAI, Google, and others are adding multimedia services for creating images, audio, or even video to their text-based chatbots, potentially displacing specialized services such as *Midjourney* or *Stable Diffusion* for images, *Speechify* or *ElevenLabs* for audio, and *Invideo AI* or *synthesia* for video.

One of the obvious areas in which generative AI tools' versatility offers benefits is in L2 writing. A tool frequently used in assessing student writing and providing means for helping students to improve their skills are automated writing evaluation systems (AWE) such as *Criterion* or *MY Access!*. Rather than using one of these (expensive) systems, written student texts can be provided to AI for analysis, with the possibility of instructing the system with detailed prompts or even an evaluative rubric. The system can also be instructed to provide individual corrective feedback on a student paper. Using a general AI tool rather than an AWE system or intelligent writing tool like *Grammarly* can provide a greater degree of customization, with the ability to focus on particular aspects of writing. Writing tools typically do not differentiate between L1 and L2 texts, but that could be done through AI, as well as specifying other program-specific priorities. An advantage of using a general AI system is that narrow AI tools are "offered in piecemeal fashion, requiring the user to navigate many distinct platforms in order to get assistance" (Tseng & Warschauer, 2023, p. 259). A further consideration is that AI tools have become more widely available and are continuously improving in functionality.

AI services seem likely to make redundant many tools and approaches to SLA through technology. Data-driven learning (DDL), for example–direct student use of corpus tools–may see through generative AI both a radical new direction in corpus use as well as an affirmation of the principles it embodies (see Crosthwaite & Baisa, 2023). Rather than relying on a prepared corpus based on collected samples that have been classified and tagged, researchers can now use already available LLMs. Applying prompts in *ChatGPT* can create the equivalent of a KWIC display (keyword in context). Other traditional instructional uses of DDL can likewise be achieved using a large language model. Existing language models can be combined with other data to create a hybrid corpus, specifically designed for language learning (see Godwin-Jones, 2021; Lee & Lee, 2024, this issue). LLMs may offer compelling new options for lexical development, for example, by extracting and analyzing examples of multiword units. Examining multiple examples of items in context can help move vocabulary learning from a literal, one-to-one correspondence (L1 to L2) to a relational understanding of words in their crucial role in collocations, set phrases/patterns, and idioms, so important for generating native-like L2 output. AI tools can supply plentiful sample sentences, but also commentary, explanations, and larger contextual passages.

One of the likely developments of interest in language learning is the option for using customized or hybrid AI systems. Bespoke "GPTs" can target specific uses, while hybrid systems can add a specially developed corpus or AI system to add language-specific functionality (grammar explanations, for example). Until recently, creating a custom AI app required programming, as discussed in the creation of a personalizable AI conversation partner in Lorentzen and Bonner (2023). That process has become much easier through services such as poe.com, that enable creation of specific AI use cases without programming, such as a grading rubric for argumentative essays (see Wang & Chan, 2023). Using that service, the prompt instructions for a particular functionality need be entered only once and become a permanent feature of the AI tool. External documents can be uploaded that specify more data or additional instruction to inform functionality. The system creates a unique URL that can then be transmitted to learners. OpenAI has enabled this ability in *ChatGPT Plus*, its subscription-based service. Lan and Chen (2024) illustrate that use with a pedagogical agent for English learners that focuses on the use of ordering and transition phrases in narratives. The ability to create such programs through written instructions, not through coding, makes their use accessible to much wider swaths of the educational community. Customized AI apps are likely to become more prevalent, following the trend in mobile computing of narrowly targeted apps.

Another option for language learners or teachers in terms of narrow AI are the many AI-powered language learning apps that have become available. That includes well-known services like *Duolingo, Babbel*, or *Memrise*, all of whom have integrated AI, as well as newer AI apps such as *Langotalk, Polyglot AI*, or *LangAI*. Some are more specific, targeting conversational abilities (*Speaking Club AI, Talkpal*) or focusing exclusively on English (*Andy English Bot*). Most of the apps require paid subscription or offer a freemium model. As with all digital tools, they vary in features and may be questionable in terms of longevity. General AI systems from OpenAI, Google, Meta, or Microsoft, stand a better chance of remaining available longer-term. The fierce competition among these companies is likely to result in ever-improving AI tools but also innovations to enable more customized AI agents.

The ever-expanding range of AI functionality is likely to mean that such systems will become, like smartphones, ubiquitous human companions. A variety of mobile apps incorporate AI in various ways, including voice chatbots based on LLMs. It seems likely that wearable and embedded devices will incorporate that functionality as well. That will come through adding an AI backend to existing voice assistants (*Alexa, Siri, Google Assistant*) as well as through entirely new apps or devices. In fact, we are already seeing the development and marketing of AI-centered devices like the AI pin, Rewind Pendant, or Rabbit AI that function as multimodal personal AI agents. In contrast to conventional voice assistants, these devices not only supply answers to queries, but can take action on behalf of the user (making travel arrangements, automatically creating summaries of meetings, suggesting dinner options, etc.). Such innovations may be of particular interest to independent language learners, as they present opportunities for integrating language use into everyday lived experiences. The wide choice of AI tools, which include general AI systems like *ChatGPT* and the customized AI tools/services/devices referenced above, may be a source of uncertainty, even anxiety in teachers and learners, but the variety of approaches does serve to enable a greater ability to match functionality to need, adding an element of user agency.

## L2 Writers' Shared Agency

For instructed language learning, AI offers many potential benefits, but also concerns, as instructors fear students will simply turn over all writing assignments to AI to complete. That may well be the case for some, but academic dishonesty is not new, although new approaches will likely be needed that are different from how cheating and plagiarism are dealt with in educational institutions. The reliance on AI detectors is not the answer, especially as they have been shown to be particularly unreliable in evaluating non-native writing (Liang et al, 2023). Some lessons can be learned from instructional practice in dealing with machine translation in instructed SLA. As discussed above, studies have shown that by and large students do not use *Google Translate* or other services to simply translate papers or assignments, but rather to check lexical choices. One of the successful strategies discussed in MT studies is to ensure a variety of assigned writing tasks, some of which specifically call for the use of MT, while others avoid its use. Research in the use of MT provide example assignments in which translation tools are integrated, such as comparing MT versions to student drafts, analyzing MT output in terms of contextual appropriateness in language register or level of formality, or engaging in extensive post-editing (see Jolley & Maimone, 2022; Vinall & Hellmich, 2022). Assignments involving MT (or AI) should be accompanied by ample reflection and discussion of experiences and results, creating greater awareness of systems' capabilities and limitations.

Similar to studies of student use of MT, research into the learner use of *ChatGPT* in L2 writing have shown that writers tend to use the tool in various stages of the writing process, rather than simply to generate a complete text to turn in (Warschauer et al., 2023). Baek et al. (2023), in a survey of student use of *ChatGPT*, found that it was most frequently used by language learners to check grammar and word selection. That study also found that students were intent on maintaining their own voice in their writing while using AI.

Of particular interest is a detailed case study of an individual student using generative AI: Jacob et al. (2023) follows the use of *ChatGPT* by a Chinese graduate student ("Kailing") in her academic writing in English. Kailing uses *ChatGPT* all through the writing process, in brainstorming ideas, doing basic research, generating sections of a first draft, and assisting in revisions. Throughout the process she uses an iterative approach to providing instructions to the chatbot, refining previous prompts. That allows her to zero in on her specific needs. She sometimes rejects AI suggestions, as they do not mesh with her own style, which is personally important for her to maintain.

Not all students are likely to imitate Kailing's approach to writing with AI. The case study does point, however, to recommendations and best practices in the use of generative AI in writing. One of those is the importance of instruction and hands-on experimentation with *prompt engineering*, a hot topic in reports of academic uses of AI chatbots. The memory capacity of *ChatGPT* allows for revising, expanding, or adjusting in other ways original prompts. Systems vary in terms of length of prompt, but extensive directions are possible in mainstream systems. Learners can, as Kailing did, reference previous output and refine the prompt towards more acceptable results. This process has been labeled "chain-of-thought" prompting (Wei et al., 2022, p. 1), and is an important component in learning to maximize the effectiveness of chatbots. That process of gradual improvement and revision in repeated exchanges with AI is also helpful in moving students away from a view of writing as transactional and towards a process orientation.

Another lesson from Kailing is the ability to interweave AI generative texts with the learner's writing and editing to achieve a personalized result. She describes her interactions with *ChatGPT* in a way that demonstrates shared agency:

> I definitely enjoyed the process more especially because of this thinking partner role of *ChatGPT*. I feel like I have like a back and forth interaction with someone who's always there. And like, and because of this interaction, I can gradually build upon my previous ideas based on my thinking and *ChatGPT* input and gradually improve the idea (Jacob et al., 2023, p. 14).

Kailing considered AI an "intellectual collaborator" (Jacob et al., 2023, p. 14), which seems a good descriptor of the authoring partnership between AI and human writer. In another study of AI assistance in L2 writing, Guo et al. (2022) found that chat sessions with an AI chatbot helped L2 writers develop persuasive reasoning in drafting argumentative essays, thus substituting for the social scaffolding offered by peer interactions. AI is able to provide feedback throughout the writing process, inviting students to engage in repeated reflection on their thinking and writing. That was a finding in Liu et al. (2021) which used a reflective thinking promotion strategy which enhanced students' writing performance and improved their self-efficacy and self-regulated learning.

In an AI world, originality and attribution are likely to be complicated, with contributions shared between AI and humans. The option to pass off to AI systems some aspects of writing, such as grammatical correctness, can allow human writers to focus on higher-level concerns such as creativity, personal voice, and reflection. That is likely to be incorporated into assessment strategies both in L1 and L2 writing. While the mechanics of writing can be assisted through AI, its social communicative function will likely need to be addressed through tasks that involve human-to-human communication. A comprehensive and balanced L2 writing program will likely include collaborative activities such as virtual exchange or peer editing. Such tasks serve the goal of moving away from instrumental and model-based writing so pervasive in our era of high-stakes testing and essay writing (McKnight, 2021).

Collaborative activities involving negotiation of meaning and finding linguistic and cultural common ground necessitate developing pragmatic and strategic competencies, that is both the ability to adjust language to social contexts and to find means to communicate when utterances are not understood. These abilities are more likely to be developed in real-world exchanges than in chatbot conversations. In addition, virtual exchange or participation in online affinity groups (such as online gaming communities) offers the opportunity for serendipitous encounters, potentially expanding learners' cultural and social horizons. In the process, misunderstandings and pragmatic errors may arise, resulting in *critical incidents* that lead to memorable linguistic learning experiences and transformative change (Canagarajah, 2021; Leaver et al., 2021). The importance of not shying away from potential conflict or destabilizing experiences in L2 encounters online is an essential element in Levine's (2020) concept of a "human ecological approach to language pedagogy" (p. 45), arguing that conflict is an inevitable aspect of human civilization, so that designing conflict avoidance keeps exchanges at an artificial and superficial level (see also Godwin-Jones, 2019b). The intense human learning potential of authentic communication with real human beings is incalculable and an affordance of real-world engagement not available from synthetic digital companions.

AI tools promote a view of language as standardized, culturally homogeneous, and monolingual (Wang et al., 2021). AI-powered voice recognition may have difficulty with dialects, accented speech, and regional variability (Ramesh et al., 2023). A researcher reported that in using a voice assistant she needed to anglicize the pronunciation of Spanish names in order to be understood (Enriquez et al., 2024). This aspect of AI's linguistic abilities could be explored in instructional settings as a way to identify and discuss AI limitations; to build metalinguistic awareness; and to encourage positive attitudes towards plurilingualism and cultural diversity. Examples of the advantages of drawing specific attention to the language-specific and multilingual uses of AI can be found in studies of bilinguals' use of digital tools. Rowe (2022) invited bilingual elementary school students to experiment with using *Google Translate* to create bilingual texts, a process shown to be personally and linguistically empowering to the children. Zhang and Li (2020) had young Canadian and Chinese biliteracy learners engage in bilingual digital storytelling to encourage collaboration and creativity across languages and cultures, offering a "counter-narrative to the neoliberal applications of new media and digital literacies" (p. 563), typical of curricula in many school settings.

Storytelling represents a powerful vehicle for connecting to cultural traditions, family relationships, and language practices; it strengthens ties to communities/generations, helps build personal identities, and can help validate the use of heritage languages (Engman & Hermes, 2021). While storytelling can be deeply personal, intimately connected to humans' social worlds and cultural values, AI tools could still play a role in areas such as brainstorming and revising, while, for digital storytelling, multimodal AI could supply appropriate images or even video clips. AI use could be of interest as well in other forms of online storytelling such as fanfiction (Sauro, 2017). In pursuing storytelling activities—and in other assigned writing tasks—it is helpful to remind students that professional writers (journalists or movie script authors, for example) use digital writing tools selectively. Professional translators have long used machine translation, but always as a drafting assistant, not for finalized versions.

An integral part of AI use in L2 writing instruction can in fact be discussion of how AI plays a role in real world settings. One option for making that connection is to have students' writing extend beyond the classroom. In teaching an advanced French writing course, "Narrating the multilingual self," Blyth (2023) used AI writing tools *Pickaxe* and *Sudowrite* to help students in composing and editing texts that were then shared online. Dooly and Darwin (2022) advise using inquiry-based learning to have students investigate social issues of importance to them personally and then submit their work in publicly available sites such as *Harbingers' Magazine*, an online journal that invites young writers to contribute. Another example is using wikis (Celik & Aydin, 2016) or *Wikipedia* (King, 2015) for L2 writing development in public spheres. The collection of texts in Ryan and Kautzman (2022) explores a variety of options for creating nondisposable or *renewable* assignments, that feature real-world integration of student writing. Such assignments help L2 learners see that there is real-world applicability for what they are learning in the classroom, and that learning how to integrate digital and AI tools into their own writing will be valuable personally and professionally.

## Framing Joint Human-Tool Agency

In an AI world, language learning through social interactions, such as peer collaboration and virtual exchange, will be accompanied by the extensive affordances of AI-integrated chatbots and other digital services. Social learning and AI assistance will for many L2 learners be embedded in a formal educational setting, with a distinct curriculum and set of learning objectives. The complexity of this intertwined learning environment is best understood using an ecological framework, which takes into account the ever-changing learner-machine-world relationship. Van Lier's (2010) emphasis on the importance of context in language learning, along with sociocultural theory highlighting the centrality of peer learning (Thorne & Tasker 2011), have formed the core of an ecological perspective on SLA.

That framework has been expanded to incorporate material culture (including software tools) and embedded communication (4E cognition theory), as well as the dynamics of shared agency and distributed roles in learning activities. *Sociomaterialism* has contributed substantially to that expanded ecological framework (see Guerrettaz et al, 2021; Thorne et al., 2021). From a sociomaterial perspective, language learning represents an *entanglement* of contributors to the learning process, with agency shared among humans and nonhumans. Sociomaterialism represents "a paradigm shift away from viewing humans as autonomous agents who dominate material resources to achieve an outcome" (Guerrettaz et al., p. 9). In that way it aligns with *posthumanism*, referencing not a time after humans but rather a decentering of humans (Toohey & Smythe, 2022).

The concept of shared and distributed agency is helpful in understanding the role of AI in language learning. Rather than considering AI systems as tools used by autonomous learner-agents, that relationship can be quite a bit more complicated (as seen in Kailing, Jacob et al., 2023), more of a partnership than a simple tool use. The process of partnership carries over to the teacher-AI relationship as well, as demonstrated in Lan and Chen's (2024) development of an adaptive, intelligent tutor. In that project, the content expert (i.e., English teacher) designs the learning interactions, tests system revisions, and mentors students pedagogically and emotionally, while the AI tutor (developed as a custom GPT) has the role of "controlling learning flow, personalized content delivery, instant feedback, and performance tracking" (Lan & Chen, 2024, p. 3). Similarly, in Liu (2024; this issue), teachers find that the optimized use of AI involved a "collaborative division of labor" (p. 169). The shared responsibilities and distributed roles that characterize learner and teacher uses of AI demonstrate the potential complexity of our relationship with such systems. While we do need to consider and analyze AI tool functionality, a more holistic perspective on its interaction with human users as well as societal implications is essential. That is all the more the case given AI's emerging presence in all domains of human activity.

While ecological theories emphasize a broad consideration of all the elements and actors in learning activities, they also examine closely the dynamics of learning trajectories, which are influenced by learners' starting points, evolving relationship with learning tools and resources, and effect of external factors and social relationships. A concept helpful in framing this dynamic is that of a *relational pedagogy*, (Kern, 2018), which emphasizes the centrality of relationships and assumes a distributed agency for human learning. A relational pedagogy "aims to foster an ability to reflect on meaning-making practices broadly, but with particular emphasis on how materials and technologies interact with social worlds and individual creativity in those practices" (Kern, 2018, p. 12). That approach aligns well with a broadly semiotic understanding of human communication, which incorporates the presence of the myriad meaning-making artefacts and sign systems in our world (Godwin-Jones, 2023a; Satar et al., 2023).

The variability in learning pathways points to the fact that outcomes, particularly in the case of AI use, can vary dramatically depending on conditions of use. That of course is not unexpected when it comes to teaching and learning or to the efficacy of software in instruction. AI, however, due to its power and ubiquity, will increase substantially the uncertainty and unpredictability. At the same time, that power and ubiquity can be harnessed to guide the learning process in particular directions. Given its ability to recall previous encounters, systems like *ChatGPT* can be used for incremental, context-aware learning. Particularly important for SLA, generative AI can be built into personal, adaptive learning environments like intelligent tutors, as demonstrated in Lan and Chen (2024). In that way, AI systems can follow the direction in SLA that emphasizes personalized goals and individualized learning (Benson, 2017; Ortega, 2017; Peng et al., 2022).

There is likely to be considerable variability in individual learner interactions with AI. Individual or small group case studies will help illustrate patterns of usage. Cluster analysis has proven to be of particular usefulness (Peng et al., 2022; Warschauer et al., 2019). Woo et al. (2024; this issue) enlists cluster analysis in analyzing the effectiveness of L2 story writing across different levels of AI assistance and integration (see also Wang et al., 2023). AI can be used in such a variety of ways, that it is important to follow successful patterns of use (and less effective strategies) which take into account learner profiles such as motivation/maturity, prior learning experiences, and proficiency level. That can enable differentiated approaches to AI integration (see Liu, 2024, this issue). Looking at individual or group use of AI needs to take into consideration not just tool use, but also what other L2 tools and resources may be part of learning histories and profiles. Studies of learner trajectories using complexity theory (Larsen-Freeman, 1997; Five Graces Group, 2009; Sundqvist & Sylvén, 2016) have shown how variable L2 development is, dependent on initial conditions and a multitude of ever-changing human (e.g., family, teacher, peer) and nonhuman resources (e.g., textbook, online affinity groups, AI). Given the proliferation of L2 resources online today, predicting learning paths and outcomes has become even more uncertain (Godwin-Jones, 2018; Sockett, 2014).

Examining successful learner trajectories integrating AI does not provide a blueprint for L2 learners generally, but it can suggest pathways and resources that may be helpful, with the important caveat that, as we have seen, affordances are emergent, dependent on a specific user in a particular time and space and on the different actors, human and nonhuman, who might be of a nature and at a level that are enabling to that learner. All tools and resources, including AI, should be considered in their potential to contribute to success in language development. That success has different characteristics, linguistic of course, but also dimensions connected to other developmental goals, like self-empowerment and social transformation (Ortega, 2024). Looking at learners holistically, language development occurs within social contexts through contact with others and with nonhuman resources. Learner identities are fluid and emergent: "Agency is realized as a capacity to make choices, create new ways of being, act against the social constraints and power hegemonies of native-non-native speaker differentials" (Klimanova, 2021). That includes the ability to perceive and act on issues of social justice (Ortega, 2017).

Awareness of and actions in support of social justice have increasingly been seen as goals in SLA (Levine, 2020; Ortega, 2024). This is a reflection, on the one hand, of social and political developments (e.g., growth of populism and nationalism, increase in discrimination and hate speech online, rising wealth inequality) and, on the other hand, the fact that language is essentially a social and cultural vehicle through which efforts at the micro level can have the potential to affect change at the meso and macro levels (Levine, 2020). Promoting social justice in SLA has led to a focus on marginalized communities, including disadvantaged minorities, migrant groups, and populations in less industrialized nations. Social norms, political agendas, and economic factors tend to favor majority languages and disadvantage or even disenfranchise speakers of other languages. Native speakerism and the overpowering presence of English as a lingua franca have been recognized as issues within research in applied linguistics (Ortega, 2017) and in CALL (Buendgens-Kosten, 2020; Sauro, 2016).

Students in the United States from diverse linguistic and cultural backgrounds have been shown to be disadvantaged in educational environments, including in computing literacy (see Jacob et al., 2024). The move towards equitable multilingualism (Ortega, 2019, 2024) is validated through the emphasis on inclusion and diversity in Indigenous communities (Canagarajah, 2021, 2023). Sociomaterialist, posthumanist, and translingual theorists have acknowledged their connections and indebtedness to concepts and traditions emanating from Indigenous communities (Canagarajah, 2023; Pennycook, 2020; Toohey & Smythe, 2022). The ways of viewing the world from a non-Western perspective not only embrace multilingualism and transculturalality, but also see a symmetry between humans and entities in nature. This ontology of nondualism and inclusivity can be helpful in considering human-machine relationships, particularly as represented in AI.

Canagarajah (2023) discusses another characteristic of Indigenous cultures that can be beneficial in developing mindsets that interact more flexibly with AI, namely the tendency to accept, even expect, ambiguity, uncertainty, and mystery in lived human experiences. That is helpful in dealing with the black box of generative AI, a system whose capabilities even its designers do not fully understand. Interacting with AI is in some ways a step into the dark, with unpredictable results. Prompt *engineering* is a serious misnomer, as the process is closer to art than science, almost like uttering an incantation in which the wording must be just right to obtain the desired outcome. According to Bearman and Ajjawi (2023), "pedagogy for an AI-mediated world involves learning to work with opaque, partial and ambiguous situations, which reflect the entangled relationships between people and technologies" (p. 1160). An acceptance of the fact that aspects of AI are ultimately unknowable may be difficult (particularly for academics) but can be helpful in learning to live with AI as a potential learning and teaching partner.

An Indigenous perspective on artefacts and tools, whether they be natural or man-made, tends to be quite distinct from the typical attitudes in Western traditions. From a Western perspective, tools are impersonal and discardable, whereas Indigenous cultures highlight "the materiality and complexity of all tools, digital and otherwise" (McKnight & Shipp, 2024, p. 5). In First Nation communities, humans and nonhuman entities are seen as intimately intertwined, not viewed within a hierarchy with humans at the apex. Indigenous cultures respect and listen to nonhuman entities. That is literally the case in a study of young bilingual Ojibwe learners in which objects in nature play a central role in a language lesson conducted in the woods (Engman & Hermes, 2021). Stretches of silence, representing the contribution of natural elements, were afforded conversational turns (recorded in transcripts within the multimodal conversation analysis). That study demonstrates that Ojibwe itself has a connection to the values and beliefs of the culture. When participant interactions took place in English, there was a dichotomy between human and nonhumans, whereas talk in Ojibwe treated nonhumans as agentive.

In Western cultures, particular objects might be seen as intimately connected to humans, especially in the hands of expert users, such as musical instruments used by virtuosos like Yo-Yo Ma or Jimi Hendrix. A high degree of intimacy — without the need for particular expertise — can be seen in the use of smartphones. For many users, smartphones have become something like a digital appendage, a constant companion used for all manners of social and practical needs and interactions. The individualization possible in smartphones provides a personal agency new in human-machine relationships (Eilola & Lilja, 2021; Godwin-Jones, 2017). As we have seen, generative AI, like smartphones, represents a shared agency between user and digital tool. As AI becomes embedded in more apps and devices, we will be conditioned, as we have been through mobile apps, to rely on AI-powered agents as a normal, integral part of our everyday existence, at school, at work, at leisure. That may even lead to emotional attachment, as demonstrated in the strong adverse reaction by users of the personality-driven chatbot *Replika* when intimacy filters were instituted (Bastian, 2023). User-tool relations are highly variable, ranging from impersonal to intimate, with many gradients in between. That applies to AI, in which the relationship is shaped by the affordances of the tool, but even more by the individual characteristics of the learner's personal L2 developmental trajectory.

## Learner Agency Through AI is Tempered by Ethical Issues and Practical Concerns

The emerging Internet of Things will likely result in AI agents embedded in everyday objects and wearable devices. That *ambient intelligence* (Dunne et al., 2022) may well supply a new dynamic for informal language learning. That vision of the future, however, needs to be seriously qualified, as this scenario is exclusive to those privileged socially and economically. Given its ease of use, multi-language support, and wide availability on mobile and tethered devices, we may view AI tools as a leveler of social and linguistic inequality. It certainly is the case that for less proficient writers of English, AI can offer a valuable lifeline, for example, in professional academic or scientific writing, where publication in English is often expected. On the other hand, for learners, a certain level of proficiency is needed to use AI tools effectively (Warschauer et al., 2023). In many parts of the world, AI and other digital tools are likely only available in school settings. McKnight and Shipp (2024) list what that can entail:

> It is important to remember what is taken for granted in opening up *ChatGPT* and putting in a writing
> prompt. This requires, in most educational instances, (transport to) a school, a secure building, a

> classroom, electrical infrastructure and reliable supply, lighting, hardware including hard drives, cables, plugs, monitors, keyboards, mice, software licences, software such as operating systems, accounts, internet access and bandwidth (p. 9)

For extramural users, infrastructure, family/community dynamics, living conditions, and available time/space are all factors that may limit non-formal access to digital tools such as AI. The principal producers of AI systems are large, for-profit companies, which reserve use of the highest performing LLMs to paying customers, an expense well beyond the reach of most of the world's population. The need for AI companies to be profitable may disincentivize them from investing in the cost of integrating minority languages (Ramesh et al., 2023). The digital divide is not disappearing (Warschauer & Tate, 2017) and AI may in fact increase disparities.

Given its power and versatility, AI enthusiasm may lead to unrealistic expectations in AI's role in language learning. More practical experience with AI may bring disillusionment. McKnight (2021), for example, initially advised students to use AI for brainstorming, but has since found that "using generative AI in the early stages of writing actually shuts down thought and criticality" (cited in Enriquez et al., 2024, p. 12). She expresses in general worries that the human element in learning is being sidelined by AI. She cites evidence that AI writing tools advantage above all higher performing students, a position echoed in Woo et al. (2024; this issue). She also has concerns over AI-powered personalized learning systems which are being developed as "cradle-to-grave learning bots" that do not recognize a learner as a person, but rather as "a set of continuously evolving data points" (cited in Enriquez et al., 2024, p. 8). The learning bots, she fears, will be integrated into surveillance mechanisms like eye tracking monitors and other biometric data to monitor learner activity. A disturbing example of that danger are the electronic headbands worn by Chinese children in an early AI school experiment (Wang et. al, 2019). LaFrance (2023) cautions that the AI systems themselves may be used to gather information about our activities, masquerading as "AI friends": "These companions will not only be built to surveil the humans who use them; they will be tied inexorably to commerce" (p. 7). Tech and media companies do not have a sterling record when it comes to user privacy–or working towards the common good–so that caution is likely justified.

A host of other issues have been raised concerning AI use generally. Those include the questionable ethics of how large language models are created, their immense energy consumption, and how training of AI is farmed out to mostly brown and black low-pay workers in lesser-developed economies (McKnight & Shipp, 2024). The data used to create an LLM may not respect creators' rights or legal permissions, a growing source of complaints from content holders and creators. The ability of multimodal AI to manufacture digital versions of human likenesses and voices ("deep fakes"), often indistinguishable from the originals, is an issue of growing seriousness. That concern was a major reason for the extended labor strike of actors in Hollywood in 2023, as they saw the potential of their identities being used, potentially without permission (or pay). Faked identities and synthetically duplicated voices have led to the spread of dis- and misinformation online. Despite efforts by governments and tech/media companies to counter AI-generated falsity and theft, the problem is likely to remain and grow, as technical progress in voice synthesis and video manipulation advances inexorably.

A major concern for implementation of AI in education is the built-in cultural biases of AI. Data collected to create an LLM is taken (without permission) from sources mostly associated with white, Western culture. It has been shown that as a result output often demonstrates a bias towards "WEIRD" perspective on the world (Western, Educated, Industrialized, Rich, Democratic; Atari et al., 2023), leaving out the perspectives of minority groups, women, and underprivileged communities generally. AI systems "reinforce and reproduce the social and racial hierarchies observed in society" (Ramesh et al, 2023, p. 1). Even when representing non-Western cultures, AI will do so from a white, Western perspective (Atari et al., 2023). That fact validates the reluctance of some First Nation cultures to use AI or to have Indigenous languages incorporated into AI data, as they by and large are not currently (McKnight & Shipp, 2024).

Language educators, lured by the potential of AI to take on different roles, may not take into consideration the ethical issues in doing so. McKnight and Shipp (2024) cite a panel presenter who instructed *ChatGPT* to take on the identity of a well-known African American author and to speak with her voice in a dialogue with the audience. That process involved "deeply problematic cultural appropriation, identity theft, potential copyright infringement" (McKnight & Shipp, 2024, p. 10). That lecturer is surely not alone in using in unreflective ways the remarkable capabilities of AI systems. One study of the use of AI in language learning described AI as being "essential for promoting cultural sensitivity, intercultural competency, and global awareness" (Anis, 2023, p. 64). In fact, these systems are far from encouraging intercultural understanding, as "generative AI privileges and centers one way of knowing and being" (Enriquez et al., 2024, p. 7). While AI may in some cases be a supportive tool in collaborative environments, the severe limitations in its cultural orientation and the ethical concerns that may arise from its use should lead one away from the temptation to use AI as a primary vehicle for conveying linguistic or cultural competency and knowledge. Critical AI literacy, incorporating an awareness of cultural biases, should lead to a perspective on AI use that emphasizes critical thinking and qualified trust in AI output.

## Conclusion: Emergent, Shared Agency in Merged AI and Virtual Reality

SLA theory emphasizes consideration of the learner as a full-fledged, physical human being (Atkinson, 2010; Levine, 2020). That means, on the one hand, taking into account the actual life experiences of the learner and, on the other hand, viewing learning not just as a brain activity but rather as embodied and enacted. This complicates considerably the conception of human communication, moving away from an exclusively cognitive, essentially verbal process. From this perspective, learner agency in SLA is multifaceted, as Witte (2023) summarizes:

> [L2 education] must also take account of the agency of the individual learner as an (inter)acting, feeling, thinking, and self-reflexive person entangled in dynamic and multilayered structures of social relations and cultural patterns, traditions, and experiences in which they continuously seek to evolve and find expression by attuning semiotically, affectively, and corporeally to ecological affordances (p. 695).

This complexity challenges, as we have seen, the ability of AI systems to develop a full communicative ability comparable to that in humans. As generative AI becomes more fully multimodal through enhanced computer vision and machine learning applied to the analysis of visual and auditory data, LLMs will become more capable of understanding embodied meanings in images and videos. That presents the possibility of developing a greater ability of extracting meaning from representations of human gestures, body movements, facial expressions, and paralanguage. That capability could in turn be integrated into immersive environments, in which the presence of nonverbal cues can create a more complete version of human communication. Through the use of realistic avatars in simulated environments, AI-enhanced virtual reality (VR) holds the potential to become a language learning partner with considerable ability to mimic a full repertoire of semiotic resources.

Most VR apps for language learning are far from fulfilling the enormous potential of VR for collaborative learning and cultural immersion (Godwin-Jones, 2023b). They do not take advantage of body/limb movements or facial recognition features in advanced avatars, nor integrate gestures in a meaningful way. As we have seen, there has been recently in SLA theory a growing recognition of the wide array of semiotic resources that go into meaning making in conversational exchanges (Klimanova & Lomicka, 2023). That includes the human body as well as the setting in which a conversation takes place. VR is able to recreate particular scenes very effectively through the use of 360-degree video and high-quality animations. These "mirror worlds" through VR (Kelly, 2019, para 1) have been shown to be effective in language and cultural learning, for example, through global simulations, as in the Parisian neighborhood project in Mills et al. (2020). Some commercial VR apps, such as *ImmerseMe*, offer simulated environments from particular cultures. Those tend to be tourist-oriented situations. Just as settings tend to be fixed in VR apps, verbal exchanges are mostly scripted role-play scenarios. User response tends to be constrained, typically limited to multiple choice selections to respond to questions or actions. That could all change through generative AI. Voice chatbots have already begun integrating AI. Added to VR as a companion, tutor, or game character, an AI bot could provide unscripted interactions with the user.

*ChatGPT* has shown that it can quite effectively take on a particular identity, real or imagined, and tailor output accordingly. Additionally, AI is fully capable of creating images or video as directed by the user. If the metaverse is to arrive one day, a major feature will be AI-created immersive environments (Jeon et al., 2023). At the same time, we need to be conscious of the fact that if AI requires a certain degree of privilege, that is even more the case for VR, as outlined in Enriquez et al. (2024):

> If you go back and think about all the access issues – all the hardware, all the software, all the subscriptions and the licenses you need to drive that VR experience – it's not inclusive at all. Yet, you have this rhetoric of "People in remote areas will be able to experience things in these immersive worlds." Well, those are often the people that don't have the digital access anyway. They're least likely to have a set of VR headsets hanging about in a storeroom down the hall when they don't have enough keyboards for kids (p. 13).

Access to tech tools is such a given in most Western settings that advocates often lose sight of the reality in other locations. Mixed reality devices, such as smart glasses, able to engage in both VR and AR (augmented reality) may at some point become both more easily wearable and affordable, but we are not there yet. In addition, privacy issues and practical/economic hurdles, as well as technical problems (compatibility across devices, for example) are likely to severely limit wide-spread implementation in educational settings.

While the feasibility of the normalized use of AI-powered VR in underprivileged communities can be ruled out for the foreseeable future, inventive educators have found workarounds. That is the case in the Virtual Tabadul project (Baralt et al., 2022) that used inexpensive Google Cardboard VR viewers in a virtual exchange involving English and Arabic learners in the US and Northern Africa. Another solution to tech use in an under resourced environment is demonstrated in an elementary school project in India (Harshitha, 2023) in which a custom chatbot was created, designed by the students themselves (representing a number of different L1s). As the school owned a single computer, the students collaborated first in the L2 (English) and in their home languages, writing down the requirements they wanted on paper, then discussing them in plenum. Then a student was assigned to use the school computer to create the English language chatbot, based on the consensus outcomes of the student discussions.

Questions of access to technology and social inequity feature prominently in a vision of the future combining AI and VR, portrayed in Neal Stevenson's, *The Diamond Age: Or, A Young Lady's Illustrated Primer* (1995). The novel presents an intriguing example of emergent human agency enabled through merged AI and immersive technologies. The protagonist, Nell, a young girl from a disadvantaged background, is given an interactive *Primer* that functions as a kind of magical, talking and animated book, pulling its reader into an imagined parallel world in which Nell acts out different roles, principally as Princess Nell. Using the book on a daily basis, she is guided through interactive storytelling and gaming scenarios to develop logical reasoning and problem solving. Nell combines those skills with her own innate traits of initiative, loyalty, and persistence to develop confidence and achieve control over her life. Although initially designed for a socially elevated family, the *Primer* is eventually mass-produced to educate poor and abandoned Chinese girls. The novel was written, according to the author, inspired by the technical optimism of the mid 90s' emerging Internet (Wong, 2024). In our era of online disinformation and hate speech, it remains to be seen whether AI will represent the empowering affordance represented in the novel.

For language learning, AI-powered VR opens up the possibility of users creating a customized immersive environment, as well as potentially L2 learning or gaming storylines. An adventure game or quest could be designed and created (through prompts) in which the L2 user accompanies a favorite action hero or historical figure. This could represent a powerful opportunity for informal language learning, as it would integrate incidental learning into an entertainment environment, making it more likely that a learner would spend the time needed for L2 learning (Godwin-Jones, 2018). Studies on informal language learning through social media, videos, or popular music have shown how effective such informal learning can be (Sockett, 2014; Sundqvist & Sylvén, 2016). Through AI, such a scenario could be highly customizable. The user could stipulate, for example, whether multiple languages would be in play or whether explicit feedback or learning reinforcement would be integrated. There would also be the potential to integrate particular linguistic or cultural areas of study or interest, which might align with a formal learning environment, if that is present. Such a complex learning dynamic clearly aligns with ecological perspectives on language learning, which emphasizes the dynamic roles of humans and nonhuman actors as well as a shared, albeit unsymmetrical agency.

Creating a customized immersive simulation as described above would until recently require many hours of programming. Today, the scenario can be created through written prompts. That means that dedicated (expensive) coders and animators are not needed and that the user (or teacher) is in full control. Now even video and 3D images can be created through verbal prompts (or images). The fact that AI enables multimedia creation and branched gaming environments through simple language prompts has repercussions beyond language learning. It reduces the need for coding in many cases, thus calling into question career options in that field. In fact, AI is bound to bring profound changes to courses of studies in higher education. If powerful simulations can be created for language learning through AI, that could prove a game changer in many fields from history or political science to STEM (science, technology, engineering, math). What will be needed in an AI world are areas in which AI cannot rival humans, namely creativity of all kinds and a pursuit of humanitarian goals of solidarity and justice. We may see a resurgence of interest in the humanities and the arts, as a result and complement to AI's reach in most areas (LaFrance, 2023; Mollick, 2023). Those are fields which deal with human creativity, belief systems, and humans' role/place in society and in nature.

Generative AI raises a host of issues related to human agency, among them: the sources of creativity, the nature of knowledge/reasoning, the definition of originality, and our relationship to nonhuman entities (O'Gieblyn, 2021). Those are deep topics that go to the essence of the meaning of life; according to ethicist Luciano Floridi (2023a), AI "may be alarming or exciting for many, but it is undoubtedly good news for philosophers looking for work" (p. 6). The social and philosophical issues become even thornier as AI marches onward towards ever enhanced capabilities and becomes entrenched in services and devices we use every day. Greater social acceptance of more AI presence in our lives could be aided through greater transparency in understanding how generative AI functions. The lack of transparency has led to calls for *explainable AI* to allow humans a peek into the black box AI represents, thereby providing a measure of trust (see Ali et al., 2023 for an overview). Given the commercial imperatives of the principal AI companies and the fierce competition among them, peering behind the LLM curtain seems at this point unlikely and may in fact not be feasible, given the complexity of how LLMs interpret input and generate output (O'Gieblyn, 2021).

Another direction that is being pursued currently in AI development is to render AI systems responsive to human codes of ethical behavior. Qureshi (2023) makes a persuasive case of the need for *moral AI*, that is, AI that learns to intrinsically care about human lives. In fact, the Delphi project is trying to move AI development in that direction through hybridizing LLMs to learn commonsense intelligence and intuitive reasoning with "the goal of teaching AI to behave in more inclusive, ethically informed, and socially aware manners when interacting with humans" (Choi, 2022, p. 153). Whether machines can learn morality is an open question, but, as government regulations and rules governing AI do not at this time seem promising, it may be worth the effort to move responsible AI producers to do more to protect against potential harm to humans through AI use.

Trying to follow the head-turning pace of advances in generative AI will give language professionals whiplash. Instead, we should focus our efforts on the goals SLA theory has embraced that feature building plurilingual competence and advocating for social justice and cross-cultural understanding. To the extent possible, we will want to enlist AI support in achieving those goals, despite multiple ethical concerns about its creation and use. One can hope that the power of AI will ultimately lead to reimagining education, moving towards building a "porous classroom," (Godwin-Jones, 2020, p. 1), which is open to integrating technology, building on local resources, and taking action towards improving the quality of human lives (Levin, 2020). To have AI help towards that goal, agentive literacy will be needed, empowering learners and teachers to integrate generative AI effectively and ethically into their professional and personal lives. We can be concerned over the role of AI in our lives, but we can also be hopeful that it can serve as a catalyst for reorienting education to be more equitable and effective. This is not just an academic issue; our students' livelihoods depend on being ready to function competently in an emerging AI world.